\documentstyle[12pt]{article}
\newcommand{\be}{\begin{equation}}
\newcommand{\ee}{\end{equation}}
\newcommand{\bq}{\begin{eqnarray}}
\newcommand{\eq}{\end{eqnarray}}

\begin{document}

\pagestyle{empty}
\begin{flushright}
 {UBCTP 92-27\\
ITEP - M7/92 \\
August  1992}
\end{flushright}
\vspace{0.4cm}
\begin{center}
{\large \bf CONTINUUM LIMITS OF ``INDUCED QCD": LESSONS OF THE
GAUSSIAN MODEL AT d=1 AND BEYOND
\footnote{This work is supported in part by the Natural Sciences and
Engineering Research Council of Canada }}\\ \vspace{1 cm} {\large I.I.
Kogan\footnote{Permanent Address: ITEP, 117259, Moscow, Russia.}
\footnote{Address after September 1, 1992: Physics Department,
 Jadwin Hall, Princeton University, Princeton, NJ 08544, USA},
A.Morozov$^2$, G.W. Semenoff and N. Weiss}\\
\vspace{0.4 cm}
{ Department of Physics, University of British Columbia\\
Vancouver, B.C., Canada V6T1Z1}
 \end{center}
\vspace{0.4cm}
\noindent
\newpage
\begin{center}
{\bf Abstract}
\end{center}
We analyze the scalar field sector of the Kazakov--Migdal model of
induced QCD. We present a detailed description of the simplest one
dimensional {($d$$=$$1$)} model which supports the hypothesis of wide
applicability of the mean--field approximation for the scalar fields
and the existence of critical behaviour in the model when the scalar
action is Gaussian.  Despite the ocurrence of various non--trivial
types of critical behaviour in the $d=1$ model as
$N\rightarrow\infty$, only the conventional large-$N$ limit is
relevant for its {\it continuum} limit. We also give a mean--field
analysis of the $N=2$ model in {\it any} $d$ and show that a saddle
point always exists in the region $m^2>m_{\rm crit}^2(=d)$.  In $d=1$
it exhibits critical behaviour as $m^2\rightarrow m_{\rm crit}^2$.
However when $d$$>$$1$ there is no critical behaviour unless
non--Gaussian terms are added to the scalar field action.  We argue
that similar behaviour should occur for any finite $N$ thus providing
a simple explanation of a recent result of D. Gross.  We show that
critical behaviour at $d$$>$$1$ and $m^2>m^2_{\rm crit}$ can be
obtained by adding a $logarithmic$ term to the scalar potential.  This
is equivalent to a local modification of the integration measure in
the original Kazakov--Migdal model.  Experience from previous studies
of the Generalized Kontsevich Model implies that, unlike the inclusion
of higher powers in the potential, this minor modification should not
substantially alter the behaviour of the Gaussian model.

\newpage
\pagestyle{plain}
\setcounter{page}{1}

\section{Introduction}

The Kazakov--Migdal model \cite{KM} for induced QCD (to be further
referred to as the KMM) has recently attracted much attention
\cite{M}--\cite{Kh-Ma}.  It is interesting as a tractable example of a
matrix model where the the integration over angular variables plays a
non--trivial role and as such, it is the natural next step in the
theoretical investigation of matrix models.  It is also a lattice
gauge theory and there is the possibility that, provided the
appropriate scaling limit exists, its large wave-length behaviour is
described by Yang-Mills theory.  The KMM contains two kinds of fields,
N$\times$N Hermitean matrices $\Phi$ which live on the sites of a
$d$-dimensional lattice and N$\times$N special unitary matrices $U$
which live on links.  At large distances, the latter are associated
with the gauge fields of Yang-Mills theory, while the former play an
essential role at short distances and presumably disappear from the
spectrum in the continuum limit.

The most direct approach to finding a continuum limit of the KMM is to
assume that $N$ is large and to use the mean--field approximation for
the $\Phi$-fields.  It is natural to assume that the mean field is
constant in space-time.  The first results in the framework of this
program were obtained in \cite{M}.  There an equation for the density
of eigenvalues, $\rho(\phi)$, of the master-field $\Phi$ was deduced
from a highly non--trivial saddle point equation with the help of an
especially simple subset of the Schwinger--Dyson equations. Using a
power-like ansatz, $\rho(\phi)\sim\phi^{\alpha}$, a solution of this
equation was found. However, in \cite{G} an exact solution of the
saddle point equations was found for the case when the action for the
scalar fields is quadratic. In that case the eigenvalues had a
semi-circle distribution $\pi\rho(\phi)\sim\sqrt{2\mu-\mu^2\phi^2}$.
It was further argued that for $d>1$ there is no critical behaviour
($\mu\rightarrow\infty$) if the critical point is approached from the
strong coupling phase ($m^2>m_{\rm crit}^2$). One must rather approach
the critical point from the weak coupling phase ($m^2<m_{\rm
crit}^2$).  In order to make the effective potential for the scalar
field stable in this situation one needs to introduce higher order (at
least quartic) terms in the bare action.  The resulting critical
behaviour will generically be first order though second order critical
points may be present for some special values of the parameters. The
$N=2$ version of the KMM with quartic terms in the potential has been
examined in the region $m^2$$<$$m_{\rm crit}^2$ in \cite{Goshe} where
such critical behaviour was indeed found.  Moreover the results of
computer simulations demonstrated a nice agreement between the
$N$$=$$2$ KMM and the predictions of the mean field analysis which has
no $a~priori$ reason to be reliable for $N=2$.

In this paper we shall present a more extensive analysis of the easily
solvable versions of the KMM.  We believe that this is necessary
before abandoning the simplest version of the model with a quadratic
potential for the scalar field or accepting the suggestion (implied in
\cite{Goshe} and \cite{G}) that there may be second order critical
behaviour in the {\it instability} region $m^2$$<$$m^2_{\rm crit}$.
(We shall call the model with such a quadratic potential a
``Gaussian'' model despite the fact that the potential is only
quadratic in the scalar ($\Phi$) fields. Its interaction with the
gauge fields ($U$-variables) is highly non-linear.)

We begin by analyzing the exactly solvable 1--dimensional KMM in
detail.  There we observe some amusing types of critical behaviour
which occur when $N\rightarrow\infty$ but which do not survive in the
thermodynamic limit (when the size of the system increases to
infinity). We shall also use the explicit results in d=1 to
demonstrate the applicability of the mean field approximation both for
these different kinds of critical behaviour as well as for the
simplified $N=2$ version of the KMM.  The mean--field results for the
continuum limit of the $N=2$ model are then generalized to any $d>1$.
In this case the critical behaviour {\it disappears} in the region
$m^2 \geq m^2_{\rm crit}=d$ (and survives only for $m^2 < m^2_{\rm
crit}$).  Thus it seems necessary to deal with the upside--down
harmonic oscillator potential and possibly to introduce higher order
terms in the scalar potential. We then argue that a similar conclusion
applies to all $N$$>$$2$. This provides us with a more transparent
explanation of the result of ref.\cite{G} concerning the absence of
criticality in the Gaussian model when $d$$>$$1$.  The origin of this
phenomenon is just a {\it logarithmic} increase of the effective
potential at infinity due to the Van-der-Monde determinants.  It is
possible to compensate for this growth by introducing {\it
logarithmic} terms into the bare potential.  These can be interpreted
as a change of the {\it measure} of the integration over matrices.
Changes of the measure of this kind have been investigated in the
somewhat simpler context of the generalized Kontsevich Model
\cite{GKM} in which case it is known to preserve all of the nice
features of the model with a quadratic potential \cite{CheMa}. This
may also be the case for the KMM though further investigation is
required.

Our analysis of the various types of critical behaviour (even those
irrelevant for the continuum limit but, rather, associated with the
large $N$ limit in $finite$ volume) indicates a wide applicability of
the mean--field approximation for the $\Phi$-fields.  The result for
the $N=2$ model also supports the suggestion that the critical
behaviour for $m^2\rightarrow m^2_{\rm crit} + 0$ (whenever it occurs)
is associated with a mean field $\Phi$ which is {\it large}, i.e.$\Phi
\gg 1$.  (This is not necessarily true if the critical behaviour is
associated with the tuning of higher order terms in the bare potential
for $\Phi$.)  Both of these features provide some justification for
the assumptions which were made in \cite{KMSW} in our analysis of the
correlators of physical observables in the KMM.

We begin in Section 2 with a review of the KMM.  Sections 3 and 4 are
devoted to a detailed discussion of the 1--dimensional KMM. In Section
5 we consider the $N=2$ model in any dimension and investigate its
critical behaviour in the mean field approximation.

\section{KMM: Some Generalities and the Mean--field Approximation}

The KMM contains Hermitean matrices $\Phi(x)$ and special unitary
matrices $U(x,y)$ which are defined on the sites (denoted by $x$) and
on the links (denoted by $<x,y>$) of a $d$--dimensional lattice,
respectively.  The partition function is given by
\bq
{\cal Z}_D\equiv
\int d\Phi[dU]\exp \left(-\sum_x{\rm tr}V(\Phi(x)) +
\sum_{<x,y>} {\rm tr} \Phi(x)U(x,y)\Phi(y)U^{\dagger}(x,y) \right)
\label{1}
\eq
where $[dU]$ is the invariant Haar measure for unitary matrices.  For
most of this Paper we shall consider the case where the action for the
scalar fields is quadratic i.e. $V(\Phi) = m^2\Phi^2$. Higher order
terms could be used to cut off the fluctuations of large $\Phi$-fields
near the critical point and would affect the detailed properties of
the continuum limit.  We shall argue that in $d$$=$$1$ these
additional terms are unnecessary.  We shall also argue that in
$d$$>$$1$ at least some additional logarithmic terms (which could also
be viewed as modifying the measure rather than the action in
(\ref{1})) are needed to produce the appropriate critical behaviour.

If the integral over $\Phi$ is taken first in eq.(\ref{1}) it gives
rise to an effective action for the gauge fields
\bq
S_{\rm eff}(U) = -\frac{1}{2} \sum_{\Gamma} \frac{\vert{\rm
tr}U[\Gamma]\vert^2}{m^{2L(\Gamma)}L(\Gamma)}
\label{SU}
\eq
where the sum is over all loops $\Gamma$ and $U[\Gamma]$ is the
product of $U$--matrices associated with links in $\Gamma$.  Besides
the conventional symmetry under gauge transformations,
\bq
\Phi(x)\rightarrow V(x)\Phi(x)V^{\dagger}(x)~~
,~~U(x,y)\rightarrow V(x)U(x,y)V^{\dagger}(y)
\label{GAUGE}
\eq
with $V(x)\in SU(N)$ there is also an invariance under multiplication
of link operators by an element of the center of the gauge group
\cite{KSW},
\bq
U(x,y)\rightarrow \omega(x,y)U(x,y)~~~,
\label{ZN}
\eq
where, for the U(N) group $\omega(x,y)$ is a unimodular complex number
and for the SU(N) group it is an element of $Z_N$, i.e.
$\left(\omega(x,y)\right)^N=1$.  This symmetry has important
implications for the properties of observables and their correlators
\cite{KMSW}.

In the ``naive continuum limit'' this effective theory resembles
Yang-Mills theory provided $D=4$ \cite{KM}.  However, this limit is
too naive in the sense that, unlike the conventional Wilson
formulation of lattice QCD which has a single, or finite number of
terms like
\bq
\frac{1}{g^2}{\rm tr} U(\Gamma)
\eq
in the action, the effective action (\ref{SU}) has the special
property that it is not sharply peaked in the vicinity of the trivial
configuration $U=I$ \cite{KMSW}.  This leads to significant
differences between the procedure for taking the continuum limit in
this gauge theory and in conventional lattice QCD.

Remarkably it is possible to do the gauge field integral first in
(\ref{1}).  The result is an effective action for the $\Phi$-fields,
\bq
S_{\rm eff}(\Phi) = m^2 \sum_x {\rm tr}\Phi^2(x) -
\sum_{<x,y>}\log I(\Phi(x),\Phi(y))
\label{2}
\eq
where
\bq
I(\Phi,\Psi) \equiv \int [dU] e^{{\rm tr}\Phi U\Psi U^{\dagger}}
=V_N\frac{\det e^{\phi_i\psi_j}}{\Delta(\phi)\Delta(\psi)}
\label{ITZUB}
\eq
denotes the integral over the $U$-matrix on a given link and we denote
by $V_N = {\rm Vol}(U(N))$ the volume\footnote{A simple way to derive
this formula is provided by the theory of random matrices.  The Gaussian
integral over Hermitean matrices $H$,
\bq
J_N=\int dH e^{-{\rm tr}H^2} =  \prod_{i=1}^{N} \int dH_{ii} e^{-H_{ii}^{2}}
 \prod_{i>j} \int dH_{ij}d\overline{H_{ij}}
 e^{-2|H_{ij}|^{2}} = \pi^{N^2/2}
\nonumber
\eq
can alternatively be expressed by diagonalizing $H$, integrating over
orbits of diagonal $H$ and the orthogonal polynomial method to compute
the remaining integral as
\bq
J_N={\rm Vol}[U(N)/U(1)^N]\int  \prod_{i=1}^{N} dh_i
\Delta^2(h)e^{-\sum_{i=1}^Nh_i^2}= \nonumber \\
{\rm Vol}[U(N)/U(1)^N]N!\int \prod_{i=1}^{N} dh_ie^{-\sum
h_i^2}\prod_{j=0}^{N-1}H_j^{2}(h_i)
\nonumber
\eq
The last product on the right-hand-side contains the norms of
orthogonal Hermite polynomials normalized so that \bq
H_k(h)=
 = \frac{1}{2^{k}} e^{h^{2}/2}(h - \frac{d}{dh})^{k}
 e^{-h^{2}/2} = h^k+\ldots \nonumber \eq  Then  ~
$\vert\vert H_k\vert\vert^2=\sqrt{\pi}k!/2^{k}$.   ~
The $N!$ factor in front of the integral comes from the permutations
of the eigenvalues $h_i$ while the volume factor arises from
integration over the flag manifold U(N)/U(1)$^N$ of unitary matrices
U(N) modulo the Cartan (diagonal) elements of U(N) which commute with
the matrix $H$ once it is diagonalized.  Comparison of these two
expressions for $J_N$ gives (\ref{GRPVOL}).} of the unitary group
(in the Haar measure $[dU]$):
\bq
V_N = (2\pi)^{\frac{N(N+1)}{2}}/\prod_{k=}^N k!
\label{GRPVOL}
\eq
The integral (\ref{ITZUB}) depends only on the eigenvalues of the
$\Phi$-fields, denoted here by $\phi_i$ and
\bq
\Delta(\phi)=\prod_{i<j}(\phi_i-\phi_j)
\eq
is the Van-der-Monde determinant. This explicit formula for
$I(\Phi,\Psi)$ has been known for a long time \cite{Kh-Ch}. In the
large $N$ limit the resulting effective theory for the $\Phi$-field
can be analyzed in the mean field approximation.

The main idea of ref.\cite{M} was to analyze the classical equations
of motion for the effective action $S_{\rm eff}(\Phi)$,
\bq
m\Phi(x) = \frac{1}{2} \sum_{y \in <x,y>}
\left(\frac{\partial}{\partial\Phi(x)}\right)\log I(\Phi(x),\Phi(y))
\label{3}
\eq
{\it together} with some simple Ward identities for $I(\Phi,\Psi)$,
\bq
{\rm tr} \left(\frac{\partial}{\partial\Phi}\right)^k I(\Phi,\Psi) =
{\rm tr}\Psi^k I(\Phi,\Psi).
\label{4}
\eq
It is important to note that this is {\it not} a complete set of Ward
identities.  They {\it do not} define $I(\Phi,\Psi)$ unambiguously.
For example
\bq
I_0(\Phi,\Psi)=e^{{\rm tr}\Phi\Psi}
\label{triv}
\eq
is also a solution of (\ref{4}). A complete set of identities would contain
analogues of these equations with operators like ${\rm tr} \left(
\Phi^l\left(\frac{\partial}{\partial\Phi}\right)^k\right)$ (where
$l$$\neq$$0$) on the left hand side). There may also be independent
identities with mixed $\Phi$ and $\Psi$ derivatives.  Unfortunately
most of them (with the exception of (\ref{4})) have a complicated
form.

Since the reduced set of identities (\ref{4}) is incomplete, it is not
clear whether $arbitrary$ solutions to the equations\cite{M}
\bq
{\cal R}(\lambda) = P\int_{-\infty}^{+\infty}
\frac{d\nu}{2\pi i} \log \frac{\lambda -
\frac{1}{2d}V'(\nu) - \frac{d-1}{d}{\cal R}(\nu) + i\pi\rho(\nu)}
{\lambda - \frac{1}{2d}V'(\nu) - \frac{d-1}{d}{\cal R}(\nu) -
i\pi\rho(\nu)} = \nonumber
\eq
\bq
\int_{-\infty}^{+\infty}
\frac{d\nu}{\pi} \arctan \frac{\pi\rho(\nu)}
{\lambda - \frac{1}{2d}V'(\nu) - \frac{d-1}{d}{\cal R}(\nu)}
\label{M}
\eq
with
\bq
{\cal R}(\mu) \equiv P\int \frac{\rho(\nu)d\nu}{\mu-\nu},
\eq
derived from them will have anything to do with the actual
KMM.\footnote{In any case it is unclear how to apply the same trick to
estimate quantities other than the eigenvalue distribution.  To study
correlation functions in the KMM one must use some more substantial
information about $I(\Phi,\Psi)$. The feature (\ref{4}) is
insufficient (see \cite{KMSW}).} While the results of Migdal and Gross
might still be relevant to the KMM for some less obvious reason this
reason needs to be clearly identified.  It was shown in \cite{G} that
(\ref{M}) posses a solution for $\rho(\mu,d)$ in the form of a
semi-circle distribution
\bq
\pi\rho(\mu, d) =\sqrt{2\mu(d)-\mu^2(d)\phi^2}
\label{cemic}
\eq
which is usually characteristic of matrix models with purely quadratic
potentials. This is not surprising since the information input in the
derivation of (\ref{M}) does not distinguish between the integral
$I(\Phi,\Psi)$ in (\ref{ITZUB}) and the trivial quadratic potential
$I_0(\Phi,\Psi)$ in (\ref{triv}). One reason why (\ref{M}) may still
be relevant is that for large $\phi_i$ (see (\ref{IZ1})),
$I(\Phi,\Psi)$ is not very different from $I_0(\Phi,\Psi)$.

The explicit expression
\bq
\mu(d)=\frac{d}{2d-1}\left[m^2(d-1) \pm \sqrt{d^{2}(m^4-1) +
(d-1)^{2}}\right]
\eq
was derived in \cite{G} by substituting the semi-circle ansatz
(\ref{cemic}) into (\ref{M}).\footnote{Here we shall use our notation,
 for example our $m^{2}$ is two times smaller then in \cite{G}}
 This expression shows that while for
$d=1$ $\mu$ goes like $\sqrt{m^2-m^2_{\rm crit}}$ in the vicinity of
the critical point this is no longer true for $d>1$ in which case the
branch of $\mu$ which is positive in the region $m^2$$>$$m_{\rm
crit}^2=d$ does not vanish when $m^2$$=$$d$.  We shall see in Section
5 below that this result has a very simple explanation and we suggest
a way to restore the critical behaviour for $d$$>$$1$.

\section{Partition Functions for Various Models in One Dimension}

In one dimension the KMM can be solved using elementary techniques.
It essentially reduces to the well known problem of a matrix-valued
harmonic oscillator.  This problem has recently been reviewed in the
context of matrix models in the recent papers \cite{BK},
\cite{CAP} as well as in \cite{KMSW} and \cite{G}.  There are as many
as four slightly different Gaussian matrix models in one dimension.
They can be defined either on an open or on a closed lattice (chain),
and they can either contain only hermitean matrices as in conventional
matrix models or they can couple to unitary matrices to form a gauge
invariant model.

We shall now list the four models and give the expression for the
partition function for each.  The relevant parameter in this Gaussian
problem is $q_{\pm} = m^2 \pm \sqrt{m^4-1}$ (rather than the ``bare
mass'' $m$). In the closed chain models $\log q$ is additive in the
length $L$ of the chain. We thus define $q_L \equiv q^L$.  In the
discussion below we use the branch $q_-$.

\subsection{Gaussian Multimatrix Model on an Open Chain}

This model contains a chain of coupled matrices with free boundary
conditions.  The partition function was computed in \cite{KMSW}:
\bq
{\cal Z}_A \equiv \int \prod_{x=1}^Ld\Phi_x e^{-m^2{\rm tr}\Phi_x^2}
\prod_{x=1}^{L-1} e^{{\rm tr}\Phi_x\Phi_{x+1}} =
\left( \frac{(1-q^2)q^L}{1-q^{2L+2}} \right)^{N^2/2}
\label{A}
\eq

\subsection{Gaussian KMM on an Open Chain}

The partition function for KMM on an {\it open chain} is essentially
the same as the partition function for the previous case.
\bq
{\cal Z}_B \equiv \int \prod_{x=1}^Ld\Phi_x e^{-m^2{\rm tr}\Phi_x^2}
\prod_{x=1}^{L-1} [dU_{x,x+1}]
e^{{\rm tr}\Phi_x U_{x,x+1}\Phi_{x+1}U^{\dagger}_{x,x+1}} =
\nonumber
\eq
\bq
= [{\rm Vol}(U(N))]^L {\cal Z}_A = V_N^L\left(
\frac{(1-q^2)q^L}{1-q^{2L+2}} \right)^{N^2/2}
\label{B}
\eq
The reason for this is that the gauge transformations (\ref{GAUGE})
can be used to substitute unit matrices $I$ for {\it all} of the
$U_{x,x+1}$ variables on the chain. The resulting integral is
identical to that of the Gaussian multimatrix model with the exception
of the factor $V_N$ which is the volume of the group U(N) with the
Haar measure [dU].

\subsection{Gaussian Multimatrix Model on a Closed Chain}

We now consider the multimatrix model on a closed chain with periodic
boundary conditions $\Phi_{L+1}=\Phi_1$.  The partition function for
this model can be easily evaluated as
\bq
{\cal Z}_C \equiv \int_{\Phi_{L+1}=\Phi_1}
\prod_{x=1}^Ld\Phi_x e^{-m^2{\rm tr}\Phi_x^2 + {\rm tr}\Phi_x\Phi_{x+1}} =
\left( \frac{q_L}{(1-q_L)^2} \right)^{N^2/2}
\label{C}
\eq

\subsection{Gaussian KMM on a Closed Chain}

For the Gaussian KMM on a closed chain the partition function is given
by \cite{CAP}:
\bq
{\cal Z}_D \equiv \int_{\Phi_{L+1}=\Phi_1}
\prod_{x=1}^Ld\Phi_x [dU_{x,x+1}] e^{(-m^2{\rm tr}\Phi_x^2 +
{\rm tr}\Phi_x U_{x,x+1}\Phi_{x+1}U^{\dagger}_{x,x+1})}~~ =
\nonumber
\eq
\bq
= V_N^L \frac{q_L^{N^2/2}}{\prod_{n=1}^N (1-q_L^n)}
\label{11}
\eq
In this case the gauge transformations (\ref{GAUGE}) can be used to
eliminate (i.e. put equal to $I$) all the $U$--matrices except for
one. This last $U$ can only be diagonalized so that $U_{ab} =
e^{i\theta_a}\delta_{ab}$. Using this idea one can derive another
expression for ${\cal Z}_D$
\cite{KMSW},\cite{CAP}:
\bq
{\cal Z}_D = (2\pi)^NV_N^{L-1}\frac{q_L^{N^2/2}}{(1-q_L)^N}
\int \prod_{a=1}^N d\theta_a \prod_{i<j}^N \left|
\frac{1~-~e^{i(\theta_a-\theta_b)}}{1~-~q_Le^{i(\theta_a-\theta_b)}}
\right|^2
\label{D2}
\eq
The right hand side of (\ref{11}) can be considered as the result of
performing the integration in (\ref{D2}).

\subsection{Comments and Applications}

We now make several observations which are important for our further
analysis of the $d$$=$$1$ model as well as for its generalization to
higher dimensions. First of all the factor
\bq
q_L^{N^2/2} = q^{LN^2/2}
\eq
is common to all four partition functions. The main difference between
them is their behaviour at the critical point $q_{\rm crit} = 1$ (i.e.
$m^2_{\rm crit} = d=1$).  Note that both the functions ${\cal Z}_A$
and ${\cal Z}_B$ for the {\it open} chain models are {\it not}
singular at this point. On the other hand the partition functions for
the {\it closed} chains do possess a singularity as $q\rightarrow 1$.
The nature of this singularity is, however, quite different in the two
models.  In the conventional multimatrix model
\bq
{\cal Z}_C \sim (1-q)^{-N^2}~~~~{\rm as}~q\rightarrow 1
\label{DC}
\eq
while in the KMM
\bq
{\cal Z}_D \sim (1-q)^{-N}~~~~{\rm as}~q\rightarrow 1
\label{DD}
\eq
The presence of this singularity reflects the occurence of zero modes
in the integral as $m^2\rightarrow d$$=$$1$. In the model $C$, $any$
constant ($x$-independent) matrix $\Phi(x) = \Phi$ is a zero mode in
this limit. This results in the power $N^2$ in (\ref{DC}). Things are
however different for the KMM $D$. Integration over the $U$-matrices
can, and indeed {\it does}, eliminate some of the zero modes which
occur at $U=I$. (This was the essence of our argument in
\cite{KMSW} against the ``naive continuum limit'' for the $U$--variables
which would imply that the vicinity of $U=I$ gives the main
contribution to the integrals over $U$.) The key observation is that
in fact {\it not all} of the zero modes disappear. The zero modes in
Cartan subalgebra survive. Indeed if the $U$'s can be diagonalized by
a gauge transformation (\ref{GAUGE}) (as we have seen is the case in
the 1--dimensional model) then the {\it diagonal} elements of the
matrices $\Phi$ do not feel $U$'s. In fact
\bq
{\rm tr} \Phi U\Psi U^{\dagger} \rightarrow \sum_{a,b}
\Phi_{ab}\Psi_{ba} e^{i(\theta_a-\theta_b)}
\label{decouple}
\eq
so that all coupling between the diagonal elements of $\Phi$ and
$\Psi$ and the $\theta$'s disappears. This fact, combined with the
absence of coupling between the diagonal and the off--diagonal
elements of $\Phi$ and $\Psi$ in (\ref{decouple}) ensures that the
zero modes $\Phi_{aa}(x)$$=$$\Phi_{aa}$$=$const cannot be eliminated
by the $U$-integration.

We can understand how the off--diagonal zero modes are eliminated by
looking at eq.(\ref{D2}).  Note that the {\it integrand} is not
singular at $q_L=1$. The singularity coming from the action is
completely canceled by the zeros of the measure. (This is a very
unusual situation which is specific to the KMM. It can be understood
as being due to the very slow i.e. logarithmic increase of the action
near its minimum.  This can in turn be traced back to the fact that
contours of arbitrarily large length contribute to the effective
action for $U$-variables. See \cite{KMSW} for details.) The net result
is that the order of the singularity in (\ref{DD}) is just $N={\rm
rank}~U(N)$. The generalization of this result to $d$$>$$1$ is
discussed in Section 4.

We now briefly discuss the implications of this singularity. According
to \cite{KMSW} the fact that the singularity is $not$ of the power
$N^2$ in the KMM (as it was in the model $C$) implies that there is no
``naive continuum limit'' for the $U$-variables (It is still possible
that a ``less naive'' continuum limit may exist.)  Note however that
the singularity still does survive. In fact it is of order $N$ which,
although much less than $N^2$, is still much greater than 1.  In the
next section we shall argue that this fact implies that non--naive
large $N$ limits may exist in the $\Phi$-sector even for finite
volume.  The resulting models are of no interest when taking the
continuum limit since they do not survive in the limit of large
volume. They are however interesting examples of critical behaviour
and they should thus be kept in mind in future investigations. We
shall also see that these limits are consistent with the mean--field
approximation for $\Phi$ and they thus provide us with additional
material to study its features.

\section{Different Types of Critical Behaviour for the
Gaussian KMM in One Dimension}

Our goal in this section is to discuss the various kinds of critical
behaviour which are possible in $d$$=$$1$ and to study the
applicability of mean--field theory in each case.  Since we have {\it
exact} expressions for the various partition functions we can easily
test the validity of the mean--field approximation. The main effect
which results when mean--field theory is exact is the factorization
property for correlators such as
\bq
\ll \left( {\rm tr}\Phi^2(x_1)\cdot\cdot\cdot{\rm tr}
\Phi^2(x_k) \right) \gg ~= \left( \ll {\rm tr}\Phi^2 \gg
\right)^k.
\label{4.1}
\eq
These particular correlators are especially suited for our purposes
since
\bq
\ll \left( {\cal N}\sum_x{\rm tr}\Phi^2(x) \right)^k \gg ~=
\frac{1}{{\cal Z}} \left(-{\cal N}\frac{\partial}{\partial m^2}
\right)^k {\cal Z}.
\label{4.2}
\eq
Thus if the factorization property (\ref{4.1}) is exact we should
find, in the large $N$ limit, that
\bq
\frac{1}{{\cal Z}} \left(-{{\cal N}\over L}\frac{\partial}{\partial m^2}
\right)^k {\cal Z}
{}~=\left( \ll {\cal N}{\rm tr}\Phi^2 \gg
\right)^k.
\label{4.3}
\eq

Note that we have introduced a normalization factor ${\cal N}$ in the
definition of these operators. The conventional choice (corresponding
to the ``naive continuum (or large $N$) limit'' which was examined in
\cite{G}) is ${\cal N}_{ncl} = 1/N^2$.  However our experience in
\cite{KMSW} indicates that one should be very cautious when fixing the
value of ${\cal N}$. The correct choice of ${\cal N}$ can (and does)
depend on precisely how the limit is taken.

To illustrate the various possibilities for critical behaviour let us
proceed directly to the most interesting case -- the KMM on a closed
chain (model $D$ from the previous section). The partition function
${\cal Z}$ is given in eq. (\ref{11}). We begin by evaluating the
correlators (\ref{4.3}).  Since ${\cal Z}$ depends on $q_L$$=$$q^L$
and we wish to differentiate it with respect to $m^2$ we must use the
fact that
\bq
-\frac{\partial \log q_L}{L~\partial m^2} = - \frac{\partial \log
q}{\partial m^2} = \frac{1}{\sqrt{m^4-1}} =
\frac{2q}{1-q^2}
\eq
Then
\bq
\ll {\cal N}{\rm tr}\Phi^2 \gg = {\cal N} \frac{2q}{1-q^2}
\left( \frac{N^2}{2} ~+~ \sum_{n=1}^N \frac{nq_L^n}{1-q_L^n} \right)
\eq
and
\bq
\ll \left( {{\cal N}\over L}\sum_x{\rm tr}\Phi^2(x) \right)^2 \gg ~=
{\cal N}^2 \left( \frac{2q}{1-q^2} \right)^2 \left\{
\left( \frac{N^2}{2} ~+~ \sum_{n=1}^N \frac{nq_L^n}{1-q_L^n} \right)^2 ~~+
\right.
\nonumber
\eq
\bq
\left.
\sum_{n=1}^N \frac{n^2q_L^n}{(1-q_L^n)^2} ~~+ ~~ \frac{1}{L}
\frac{1+q^2}{1-q^2}
\left( \frac{N^2}{2} ~+~ \sum_{n=1}^N \frac{nq_L^n}{1-q_L^n} \right)
\right\}
\label{ea}
\eq

We are now ready to discuss the different limits.  We are concerned
with the limit $N\rightarrow \infty$, $q\rightarrow 1$ and possibly
$L\rightarrow \infty$.  Let us define $\epsilon_L$$=$$1-q_L$$=$$
1-(1-\epsilon)^L$ and $\epsilon\equiv\epsilon_1=1-q$. We shall always
be concerned with the limit $\epsilon\rightarrow 0$ so that $\epsilon$
is assumed to be small.  We shall also assume that
$\epsilon_L$$\ll$$1$ so that $\epsilon L$$\ll$$1$.

\subsection{Limit (a): $N\epsilon_L \gg 1$}

Let us first consider the case when $N\epsilon_L \gg 1$.  This is the
naive large $N$ limit (considered in \cite{G}).  In this limit we pick
up only the terms with the highest powers of $N$.  If we choose the
normalization ${\cal N} = 1/N^2$ these are the terms which are finite
in the limit $N \rightarrow \infty$.  From the above formulae we find
in this limit
\bq
 \ll \left( {{\cal N}\over L}\sum_x{\rm tr}\Phi^2(x) \right)^k \gg ~=
\left( \frac{{\cal N}N^2}{2\sqrt{m^4-1}} \right)^k
\left(1~+~{\cal O}(1/N) \right) ~=
\nonumber
\eq
\bq
\left( \ll {\cal N}{\rm tr}\Phi^2 \gg \right)^k
\left(1~+~{\cal O}(1/N) \right),
\eq
We thus see that the factorization property is valid in this limit.
Furthermore the particular value of $\ll{\cal N}{\rm tr}\Phi^2 \gg =
1/2\sqrt{m^4-1}$ as well as the other averages (see below) are
consistent with the prediction of the semicircular distribution
$\pi\rho(\phi) = \sqrt{2\epsilon-\epsilon^2\phi^2}$
\cite{G}.

\subsection{The limit (b): $N \rightarrow \infty$, $N\epsilon_L \ll 1$ }

The limit $N \rightarrow \infty$ with $N\epsilon_L \ll 1$ can be quite
different than the previous one since terms in (\ref{ea}) which have
less powers of $N$ can be more singular as $\epsilon_L \rightarrow 0$.
In fact since
\bq
\sum_{n=1}^N \frac{nq_L^n}{1-q_L^n} = \sum_{n=1}^N \frac{n}{n\epsilon_L}
\left(1~+~{\cal O}(n\epsilon_L) \right) = \frac{N}{\epsilon_L}
\left(1~+~{\cal O}(N\epsilon_L) \right),
\eq
and
\bq
\sum_{n=1}^N \frac{n^2q_L^n}{(1-q_L^n)^2} =
\frac{N}{\epsilon_L^2}\left(1~+~{\cal O}(N\epsilon_L) \right)
\eq
we have:
\bq
\ll {\cal N}{\rm tr}\Phi^2 \gg~ =
\frac{{\cal N}}{\epsilon_L} \left( \frac{N^2}{2} + \frac{N}{\epsilon_L}
\left(1~+~{\cal O}(N\epsilon_L) \right) \right) =  \nonumber
\eq
\bq
\frac{{\cal N}N}{\epsilon_L^2}
\left(1~+~{\cal O}(N\epsilon_L) \right)
\eq
and
\bq
\ll \left( {{\cal N}\over L}\sum_x{\rm tr}\Phi^2(x) \right)^2 \gg~=~
\frac{{\cal N}^2}{\epsilon_L^2} \left\{
\left( \frac{N^2}{2} + \frac{N}{\epsilon_L}
\left(1+{\cal O}(N\epsilon_L) \right) \right)^2 +
\right.
\nonumber
\eq
\bq
\left.
\frac{N}{\epsilon_L^2} \left(1+{\cal O}(N\epsilon_L) \right) +
 \frac{1}{L\epsilon_L}\left(\frac{N^2}{2}+\frac{N}{\epsilon_L}
\left(1+{\cal O}(N\epsilon_L) \right) \right)  \right\} =
\eq
\bq
\left(\frac{{\cal N}N}{\epsilon_L^2}\right)^2
\left(1~+~{\cal O}(N\epsilon_L, \frac{1}{N}, \frac{1}{NL}) \right)
\nonumber
\eq
We see that in this limit the results are somewhat different. Firstly
in order for a reasonable limit to exist ${\cal N}$ should be equal to
$1/N$ (rather than $1/N^2$). Secondly the averages behave as
$\epsilon_L^{-2k}$ (rather than $\epsilon_L^{-k}$). But despite these
differences the factorization property still remains valid for $k \ll
N$.  Note however that the relevant master field is clearly different
from that of ref.\cite{G} and there is no reason to believe that it is
described by a semicircular distribution. Our analysis so far which is
based on the calculation of the specific correlators $\ll \left(
\sum{\rm tr}\Phi^2 \right)\gg$ is not sufficient to
completely rule out this possibility.

To demonstrate the failure of the semicircular distribution, one
should evaluate the more complicated averages $\ll{\cal N}_k{\rm
tr}(\Phi^{2k})\gg$ which cannot be directly extracted from the above
partition functions. If the distribution of eigenvalues obeys the
semicircle law these averages should be equal to
\bq
\ll{\cal N}_k{\rm tr}\Phi^{2k}\gg_{\rm semicircular} =
\frac{(2k)!}{k!(k+1)!} \frac{1}{(2\epsilon)^k}
\label{seci}
\eq
Any deviation of these averages from this formula indicates that the
distribution is not semicircular.  We compute these averages below.

Before we proceed to the calculation note that the normalization
factor ${\cal N}_k$ is equal to
\bq
{\cal N}_k \equiv \frac{1}{N}(N{\cal N})^k.
\label{norm}
\eq
This will be true for all the types of limits which we consider.  The
reason for this is that every trace should be accompanied by a factor
of $1/N$ if it is to be substituted by an integral with an
eigenvalue-distribution function $\frac{1}{N}{\rm tr}...
\rightarrow \- \int d\phi\rho(\phi)...$ normalized so that $\int
d\phi\rho(\phi) = 1$. The factor $(N{\cal N})^k$ can be absorbed into
the normalization of the $\Phi$-field in which case the role of the
effective Plank constant in the integral would be played by $(N{\cal
N})/N = {\cal N}$ (the factor of N in denominator is associated with
traces in the action).  Applicability of the saddle point
approximation requires ${\cal N}$ to be small.

To simplify our evaluation of $\ll{\cal N}_k{\rm tr}\Phi^{2k}\gg$ we
shall begin with the case $L$$=$$1$. It is straightforward to
generalize to the case for general L.  Up to terms which are
negligible since they are of higher orders in 1/L and 1/N, it can be
obtained by substituting $q_L$ for $q$. The integral of interest
\cite{KMSW}
\bq
\int d\theta_a \Delta^2(e^{i\theta_a})
\int d\phi_{ab} \exp\bigl(-\sum_{a,b}\vert\phi_{ab}\vert^2
(m^2-\cos\theta_{ab})\bigr)~{\rm tr}\Phi^{2k}
\label{de1}
\eq
can be greatly simplified in the vicinity of the critical point (This
integral should, of course, be divided by a similar integral with
$k$$=$$0$ in order to give $\ll{\cal N}_k{\rm tr}\Phi^{2k}\gg$).
Indeed, for $k$$=$$0$ this integral is just equal to (\ref{DD}), and
the integrand is non--singular at $q$$=$$1$. At the limiting point
$q$$=$$1$ the integral is simply equal to unity.  It follows that the
measure of integration over $d\theta$ in (\ref{de1}) becomes trivial
at the critical point.  Thus in order to compute our correlators we
need only evaluate the contributions of propagators
\bq
\ll\phi_{ab}\phi_{cd}\gg~=
\bigl(2(m^2-\cos\theta_{ab})\bigr)^{-1}\delta_{ad}\delta_{bc}
\eq
Since we are interested in the most singular terms at small $\epsilon
= 1-q\sim \sqrt{2(m^2-1)}$, we can take all the differences
$\theta_{ab} = \theta_a - \theta_b$ in the propagators to be small, so
that
\bq
\ll\phi_{ab}\phi_{cd}\gg~\rightarrow
\bigl(\theta^2_{ab}+ 2(m^2-1)\bigr)^{-1}\delta_{ad}\delta_{bc}
\eq
The propagator for the {\it diagonal} components of $\Phi$ is
independent of $\theta$ and equals $1/\epsilon^2$. The contribution of
any off-diagonal component
\bq
\sim \int d\theta_{ab} \bigl(\theta^2_{ab}+ \epsilon^2\bigr)^{-1}/
\int d\theta_{ab}  = (\pi/2\epsilon)(\pi)^{-1} = 1/2\epsilon
\eq
It remains to compute the number of diagonal and off--diagonal
components arising from ${\rm tr}\Phi^{2k}$.  Their contributions are
easily distinguished since those of the diagonal components give rise
to higher powers of $1/\epsilon$ but to lower powers of $N$. There are
thus $N$ diagonal components and $N^2(1+{\cal O}(1/N))$ off--diagonal
ones. Let us first consider the limit (a) where we retain only the
terms of order $N^2$.  Now, since $k \ll N$ we can assume that
integrations over different off--diagonal elements are independent.
Using a combinatorial argument (see below, this gives the contribution
of the off-diagonal components to the correlator as:
\bq
{\cal N}_k s_k^{0} N^{k+1}/ (2\epsilon)^k
\eq
with $s_{k}^{0} = (2k)!/k!(k+1)!$.
With the proper choice of normalization factor,
\bq
{\cal N}=1/N^2
\eq
thus
\bq
{\cal N}_k = 1/N^{k+1}
\eq
this coincides with (\ref{seci}), and confirms the result of \cite{G}
for the limit (a) in one dimension.

In the limit (b) we instead take into account the contribution of only
the diagonal elements of $\Phi$.  In contrast to the off-diagonal
case, the contribution to ${\rm tr }\Phi^{2k}$ is given by the N times
the average of one component to the power 2k.  This gives a factor of
$N$ (in accordance with the proper normalization ${\cal N}=1/N$ in
this case) times
\bq
\int d\phi (\phi)^{2k}e^{-\epsilon^2\phi^2/2}
\sim (2k-1)!!\epsilon^{-2k} = s_{k}^{k}\epsilon^{-2k}
\eq
In this limit we clearly get a distribution function of the form
\bq
\rho(\phi) = \frac{\epsilon}{\sqrt{2\pi}}
e^{-\epsilon^2\phi^2/2}
\eq
rather than the semicircular distribution.  Note that this
$\rho(\phi)$ is non-vanishing on the entire axis.

\subsection{Limit (c): $N\epsilon_L \sim 1$}

We shall now consider the limit where $N$ is large, $\epsilon$ is
small but $N\epsilon$ is of order 1.  This is a kind of a ``double
scaling'' limit somewhat different from the type usually encountered
in matrix models of gravity where $N$ and $L$ are correlated and $N$
can be interpreted as a new continuum variable giving rise to a new
space--time dimension $d\rightarrow d+1$.  The formulae for the
correlators in this limit are
\bq
\ll{\cal N}_k {\rm tr}\Phi^{2k} \gg~= (N{\cal N})^k
\sum_{j=0}^{k} s_K^{(j)}
\frac{(N/2)^{k-j}}{\epsilon^{k+j}}(1~+~{\cal O}(1/N)).
\eq
In the limits (a) and (b) the first ($j=0$) and the last ($j=k$) terms
in this sum are dominating respectfully. The coefficients $s_k^{(j)}$
can be defined from the following combinatorial problem. Take ${\rm
tr}\phi^{2k} = \sum_{\{a_i\}}\phi_{a_1a_2}...\phi_{a_{2k}a_1}$ and
consider all the pair contractions of $\Phi$'s (Wick rule), each
contraction being $<\phi_{ab}\phi_{cd}>
\equiv\delta_{ad}\delta_{bc}(\alpha + \beta\delta_{ab})$, where
$\alpha = 1/2\epsilon$ and $\beta=1/\epsilon^2$.  The $s_k^{(j)}$ is
by definition the leading term (with the highest possible power of
$N$) in the coefficient in front of $\alpha^{k-j}\beta^j$. Let us draw
a picture (Fig.1a) where dots correspond to the $\Phi$-matrices. Every
solid-line contraction (Fig.1b) of dots, gives rise to a factor of
$\alpha$ and makes an identification $a=d, b=c$; while every wavy-line
contraction (Fig.1c) contributes a factor of $\beta$ and identifies
$a=b=c=d$. One can easily check that any $intersection$ of two solid
lines ``eats up'' a factor of $N^2$, while any intersection of a solid
line and a wavy line eliminates one $N$. Thus the leading-$N$
contribution arises in the ``planar'' limit, when the only lines of
contraction which are allowed to intersect are wavy lines. Thus,
estimation of $s_k^{(j)}$ is a clear combinatorial problem of
enumeration of pair contractions of the $2k$ points on a circle by
$k-j$ solid and $j$ wavy lines, such that only wavy lines are allowed
to intersect.  One can easily derive a recurrent relation:
\bq
s_{k+1}^{(i)} = \frac{k+1}{k+1-i} \sum_{j=0}^i \sum_{l=j}^{K-i+j}
s_l^{(j)}s_{k-l}^{(i-j)};~~~~~~i \leq k.
\label{rec}
\eq
It is enough to note, that the first solid line separates the circle
into two new circles of the lengths $l$ and $k-l$. The factor
$\frac{k+1}{k+1-i}$ arises because there are $2(k+1)$ points to which
the first solid line can be attached, while at the end there are as
many as $2(k+1-i)$ points at the ends of all solid lines. The
relations (\ref{rec}) should be supplemented by ``initial
conditions'': $ ~s_0^{(0)}=1;~~~s_k^{(k)} =(2k-1)!! $, the first one
is obvious, the second is just the number of $all$ possible
(intersections allowed) contractions of $2k$ dots by $k$ wavy lines.
This information is enough to find any $s_k^{(i)}$.
In particular,
\bq
s_k^{(0)}=2^k\frac{(2k-1)!!}{(k+1)!}=\frac{(2k)!}{k!(k+1)!}
\eq
and
\bq
s_k^{(i)}= s_{k}^{0}\frac{k!}{(k-i)!i!}\frac{P_i[k]}{(k+2)\ldots (k+i)}
 =\frac{(2k)!}{(k-i)!(k+i)!i!}P_i[k]
\eq
where $P_i[x]$ are certain polynomials of degree i-1:
\bq
P_0=\frac{1}{x+1},~~~P_{1}=1,~~~P_{2}=x+1, \nonumber\\
P_{3}=x^2+11x+48,~~~P_{4}=x^3+21x^2+218x+1248,\ldots
\eq

\subsection{Limit (d): The Modular form}

Finally there seems to be one more non--trivial (and, perhaps,
attractive) possibility. It comes to mind when one looks at the
formula (\ref{11}) for ${\cal Z}_D$. Clearly there should exist a
large $N$ limit, when
\bq
{\cal Z}_D \rightarrow \frac{q_L^s}{\eta(q_L)}
\label{eta}
\eq
Naively $s = N^2/2 + 1/24$, but one can think about reinterpretation
of the $U(N)$-dimension $N^2$ as, say,
\bq
N^2 = N + 2\frac{N(N-1)}{2} = \sum_{n=1}^N 1 + 2\sum_{n=1}^{N-1}n
\stackrel{N \rightarrow \infty}{\rightarrow} -\frac{1}{2} + 2
\left(- \frac{1}{12} \right),
\eq
i.e. as being regularized with the $\zeta$-function technique, thus
giving $s$ in (\ref{eta}) some finite value ($s = -1/8$).  More
interesting, the elliptic function $\eta(q) ~\equiv ~~\-
q^{1/24}/\prod_{n=1}^{\infty}(1-q^n)$ in the denominator in
(\ref{eta}) has an essential singularity as $q \rightarrow 1$:
\bq
\frac{1}{\eta(q_L)} ~\sim ~ \sqrt{(1-q_L)} \exp \frac{\pi^2/6}{1-q_L}.
\eq
This gives rise to expressions for the correlators like
\bq
\ll \left( {\rm tr}\Phi^2 \right)^k \gg = \left( \frac{\partial\log q}
{\partial m^2} \frac{\pi^2/6}{(1-q_L)^2}\right)^k(1 + {\cal
O}(1-q_L)),
\eq
which again obey the factorization property in the leading
approximation.  The role of $N$ of the previous cases is now played by
$N_{eff} \sim \frac{1}{1-q_L} = \frac{1}{\epsilon_L}$, which is large
near the critical point; in this sense this limit can be just some
particular case of (c).

Note also, that $\eta(q)$ in the denominator of (\ref{eta}) is equal
to the square root of Laplace operator on $two$--dimensional surface
(with two $real$ dimensions, $q$ plays the role of the modular
parameter, describing the shape (complex structure, to be exact) of
the torus).

\subsection{Alternative Large--$N$ Limits Beyond $d=1$}

We saw in the previous Sections that the critical behaviour in a
Gaussian model arose from the presence of zero modes of the quadratic
form in the action.  We also saw that in the Gaussian d=1 model
exactly the $N$ (= rank~$U(N)$) of the $N^2$ (= dim~$U(N))$ zero modes
of the $\Phi$-field survive after the $U$--integration. Here, we shall
argue that the case of $d>1$ is much more subtle.

Let us first consider an analogue of the model $C$ from Section 3 in
$d$ dimensions. The partition function is given by
\bq
{\cal Z}_C^{(d)} = \int \prod_x d\Phi_x e^{-m^2{\rm tr}\Phi_{x}^{2}}
\prod_{<x,y>} e^{{\rm tr}\Phi_x \Phi_y},
\label{Cd}
\eq
The result of the integral is of course the determinant of the lattice
Laplace operator raised to the power $-N^2/2$. Zero modes on a lattice
without boundaries (such as the $d$--dimensional torus) are constant
($x$-independent) matrices $\Phi$. There are thus $N^2$ of them. On a
rectangular lattice they arise when the mass vanishes i.e. when the
parameter $m^2 = m_{{\rm crit}}^2 = d$.

If we proceed to the KMM (\ref{1}) the theory becomes non--linear and
the number of zero modes is no longer $N^2$. In fact when $d$$>$$1$ it
is no longer trivial to show that there are $N$ zero modes associated
with the diagonal matrices $\Phi$ (i.e. those belonging to the Cartan
subalgebra of $U(N)$).  This is already clear because diagonal
matrices are not {\it a priori} distinguished in eq.(\ref{1}). In the
one--dimensional model $D$ they were distinguished but only after the
$U$--matrices were diagonalized.  Then the diagonal components of
$\Phi$ decoupled from $U$--fields (since $\vert U_{ii} \vert^2 = 1$
for diagonal $U_{ij}$), and the $N$ corresponding zero modes survived
the integration over $U$.  Unfortunately there is no analogue of this
procedure in the higher--dimensional case. Gauge transformations
(\ref{GAUGE}) are not enough to make all the $U$'s at all the links
diagonal. (The most one can do is to eliminate all the $U$'s on any
maximal tree of the lattice and diagonalize {\it one} more $U$.) It is
however possible to diagonalize $\Phi$ at all the sites. (The
difference is that for $d$$>$$1$ there are more links than sites). As
a result the $U$--integral depends only on the eigenvalues of $\Phi$.
The resulting formula \cite{Kh-Ch} is given by
\bq
I(\Phi,\Psi) = V_N \frac{{\rm
det}_{(ij)}e^{\phi_i\psi_j}}{\Delta(\phi)\Delta(\psi)} = V_N
\frac{\sum_P (-1)^P
e^{\sum_{i=1}^N\phi_i\psi_{P(i)})}}{\Delta(\phi)\Delta(\psi)},
\label{IZ}
\eq
where $\Delta(\phi) = \prod_{i<j}^N (\phi_i-\phi_j)$ and the sum is
over all the $N!$ permutations $P$ of the $N$ variables
$\{\psi_1,\cdots,\psi_N\}$.

Despite the above argument we still claim that there are $N$ zero
modes even when $d$$>$$1$.  To understand the idea of the derivation
we begin with a simple model which imitates some of the features of
the $d$--dimensional lattice. We consider the KMM on a lattice
consisting of a single site with $d$ links attached to it. This
``lattice'' is sometimes called a ``bouquet'' and is shown in Fig. 2.
After the $U$-integrations the partition function for the KMM on this
lattice is
\bq
{\cal Z}_D[{\rm bouquet}] = \int \frac{\prod_{i=1}^N d\phi_i
e^{-m^2\phi_i^2}}{(\Delta(\phi))^{2(d-1)}} \left( \sum_P (-)^P
e^{\sum_{i=1}^N\phi_i\phi_{P(i)}} \right)^d.
\label{IZ1}
\eq
This integrand is non-singular as $\phi_i-\phi_j \rightarrow 0$ since
the order of the zeros in the numerator is more than enough to cancel
the zeros in the denominator. Zero modes of the $\Phi$-field
contribute to the integral at large $\phi_i$.

To find the dominant singularity let us simplify the problem even more
by considering first the case $N$$=$$2$. In this case the eigenvalues
$\phi$ of $\Phi$ can be written as $\phi_{1,2}
=\frac{1}{\sqrt2}(\phi_{\rm tr} \pm\varphi)$. The variable $\phi_{\rm
tr}$ which is associated with the $U(1)$ factor of $U(N)$ decouples
(for any $N$) and  gives rise (for any $N$) to a factor
\bq
\int d\phi_{\rm tr} e^{-(m^2-d)\phi_{\rm tr}^2} \sim
\frac{1}{\sqrt{N(m^2-d)}}
\eq
in the partition function. (This is why we usually have one trivial
zero mode if we consider $U(N)$ instead of the $SU(N)$.  For $N$$=$$2$
the contribution of $\varphi$ to the partition function is
\bq
{\cal Z}_D[{\rm bouquet}]\sim \int d\varphi e^{-m^2\varphi^2}
\frac{(e^{\varphi^2}-e^{-\varphi^2})^d} {\varphi^{2(d-1)}}\sim \nonumber \\
 \sum_{k=0}^d  (-1)^{d-k}\frac{d!}{k!(d-k)!}(m^2+d-2k)^{d-3/2}
\label{SU2}
\eq
The first singularity which arises as $m^2$ decreases from infinity is
at $m^2 = m_{\rm crit}^2 = d$.  Note that, unlike the situation in
$d=1$ where the partition function itself is singular at the critical
point, here it is non--analytic so that high enough orders of its
derivatives are singular.  In d=4 the only the third and higher order
derivatives by $m^2$ become singular, the standard behaviour of a
 third order phase transition.  This non--analyticity comes from the
large $\varphi$ behaviour of the integrand which contains the factor
$e^{(d-m^2)\varphi^2}$. It is associated with the limit where large
$\phi$ is damped only by a logarithmic potential in the action, rather
than a Gaussian. This is a result of the $\varphi^{2(d-1)}$ term in
the denominator which comes from the Van--der--Monde determinants.  We
clearly see how the $U$-integration smooths the singularity at
$m^2=d$: If this were a Gaussian Hermitean matrix model the
Van--der--Monde determinant would appear in the numerator rather than
the denominator and the integration would really be singular at
$m^2=d$.

It is clear that the partition function can be made more singular at
$m^2=d$ by inserting a power of $\varphi$ in the measure for the
$\Phi$ integration. This option, which in the general case amounts to
inserting a power of $\det\Phi$ in the integration measure, is
discussed in more detail in Section 5.

It is straightforward to generalize this to arbitrary $N$. The main
singularity comes from the contribution of the identity permutation
$P=I$ in the sums (\ref{IZ1}) on every link. The corresponding
contribution is $\sim (m^2-d)^{N(N-1)(d-1)/2-N/2}$. (We have include
the $U(1)$ factor of $U(N)$).  As a byproduct this calculation reveals
an interesting feature of the model. Despite the non--linearity of the
Gaussian KMM the critical value of $m^2$ is just the same as in the
quasiclassical approximation. This observation is consistent with our
hopes that this model may be exactly soluble.

Note that one should not be confused by the fact that the term $P=I$
is only one of $N!$ terms in the sum (\ref{IZ1}) and that there is
also a product over all of the links of the lattice. In fact the
situation is similar to evaluating $\bigl(1~
+~(N!\epsilon^N)^{-1}\bigr)^{\# {\rm of~links}}$ and it may seem that
the second term is negligibly small.  We saw however in the previous
section that this does not happen.  (The factor of $N!$ is hidden in
the $\prod_{n=1}^N(1-q^n) \sim N!(1-q)^N$ as $q \sim 1$).  Indeed note
that $N!\epsilon^N \sim (N\epsilon)^N$ and in the limit (b), when
$N\epsilon \ll 1$ this term obviously dominates.

Unfortunately a similar analysis of the most intriguing limit (d) is
difficult to do, even approximately, in higher dimensions.  It almost
certainly requires an exact solution of the $d$--dimensional KMM
(which may not be an absolutely hopeless problem) and it will be left
for future investigations.

\subsection{The Large Volume Limit}

To end this section we comment on the large $L$ limit and we explain
why only case (a) of the large $N$ limits considered above has
anything to do with continuum limit of the KMM.  The crucial point is
that in this limit one {\it may not} assume that $q^L$ is close to
unity ($\epsilon_L \ll 1$) whenever $q$ is near $1$ ($\epsilon \ll
1$). One can say that the infrared critical point $q=1$ is
exponentially unstable under renormalization group evolution with
increasing $L$. (Recall \cite{KMSW} that it is $q_L = M_L^2 \pm
\sqrt{M_L^4-1}$ rather than the mass $M_L$ that is renormalized in the
simple manner $q_L = q^L$.)  It thus follows that our considerations
in this section are not relevant for the large $L$ limit of the
theory. This is made even clearer by considering the free energy
\bq
-\log {\cal Z}_D = - \frac{N^2}{2}\log q_L + \sum_{n=1}^N \log
(1-q_L^n)
\label{FE1}
\eq
which in the normal thermodynamical limit should be proportional to
the size $L$ of the system as $L \rightarrow \infty$. This is
certainly true for the first term on the r.h.s. of the above equation
since $\log q_L = L\log q$ for $any$ $q$). But $\log(1-q_L^n)
\rightarrow 0$ as $L \rightarrow\infty$ unless $q$ is fine tuned with
exponential accuracy so that $1-q_L^n$, which normally $\sim(1 -
e^{-LN\epsilon})$ can be replaced by $ nL\epsilon \ll 1$. This becomes
even more spectacular in the limit (d) where the main contribution to
the free energy is
\bq
\log
\eta(q_L) \rightarrow \frac{\pi^2/6}{1-q_L} \sim \frac{\pi^2/6}
{L\epsilon}
\nonumber
\eq
which is {\bf inversely} proportional to the size $L$ of the system.
Thus for the study of the large $L$ limit only the first term in
(\ref{FE1}) is relevant. This is of course the one which is important
for the limit (a), which was examined in \cite{G} and which leads to
the semicircular distribution.

One could think that this argument is important to explain the
irrelevance of limits (b)--(d) only in one dimension. Indeed, there is
a great difference between the KMM for $d$$=$$1$ and for $d$$>$$1$.
The main feature of the KMM in dimensions greater than one is the
presence of many closed contours on the lattice while there is only
one such contour in model $D$ in one dimension.  If there were no
closed contours, all the $U$--variables could be gauged away by a
gauge transformation (\ref{GAUGE}). This very reason makes the large
$L$ limit of the KMM in one dimension (where only one matrix $U$
survives) very different. For the study of KMM in higher dimensions it
is more reasonable to look at the 1--dimensional KMM with small values
of $L \sim 1$. However, one can easily see that this argument does not
help the critical phenomena arising due to the zero-modes: all of them
become irrelevant in infinite volume in $any$ dimensions\footnote{The
only possibility for these types of critical points to survive can be
some sophisticated limits (similar to conventional double scaling
limit of the gravity-models), when $L$ and $N$ are adjusted to
increase in a correlated fashion, so that $Ne^{-L\epsilon} = const$,
or, in other words, when $N$ grows exponentially with the decrease of
the lattice spacing $a$: $\log N\sim 1/a$. This is extremely unnatural
from the point of view of QCD, where ane can rather expect that the
correlation (if any) is $N \sim 1/\log a$.}.

\section{KMM Beyond One Dimension: From the KMM to a Modified KMM}

Another simple example of the KMM which can be analized rather easily
is the model with $N=2$. We already used this example in the previous
section, as well as in our considerations of the correlators in
\cite{KMSW}. We shall now analyze the mean--field approximation in the
$\Phi$-sector of this model and see that it is, first, in agreement
with the exact solution at $d=1$, second, it does not change much when
$d$ is increased to be greater than one, and, third, implies the
existence of the nice mean--field in $d>1$ KMM. Of course these
results do not $need$ to mean too much for the large-$N$ limit of the
KMM at $d>1$, and the conclusion of ref.\cite{G} still can be
unavoidable in that situation. However, we do not see any obvious
reason for this in our simplified analysis.

The basic formula for the study of the $\Phi$-sector of KMM for $N=2$
is the integral (\ref{SU2}). While we introduced this formula for a
specific $bouquet$ lattice, it has a more universal meaning: it
describes the effective theory of a $constant$ master field on a
conventional rectangular $d$-dimensional lattice. The value of the
master field is therefore defined from the equation of motion for
effective potential
\bq
V_{\rm eff}(\varphi) = m^2\varphi^2 + (d-1)\log \varphi^2 - d\log
\sinh \varphi^2,
\label{pot}
\eq
which look like
\bq
m^2 + \frac{d-1}{\varphi_0^2} = d\coth\varphi_0^2
\label{eqm}
\eq
while
\bq
\frac{\partial^2 V(\varphi_0)}{\partial\varphi^2} =
4\varphi_0^2 \left(
\frac{d}{\sinh^2\varphi_0^2} -  \frac{d-1}{\varphi_0^4} \right).
\eq
If $m^2 < m^2_{\rm crit} = d$, this potential has a form of Fig.3a, it
usually has a minimum, but needs and needs some higher order terms to
be stabilized at infinity \cite{Goshe} (puncture line on the picture);
while in the case of $m^2 > m^2_{\rm crit} = d$ it looks like Fig.3b.
and has a clear minimum at $\varphi = \varphi_0$ and does not require
anything in order to be stable.

In the case of $d=1$ the logarithmic term in (\ref{eqm}) disappears
and one can get:
\bq
\varphi_0^2 = \frac{1}{2} \log \frac{m^2+1}{m^2-1}
\eq
and
\bq
{\cal Z}_{mean~field} \equiv e^{-V_{\rm eff}(\varphi_0)}
{\rm det}^{-1/2}\frac{\partial^2V(\varphi_0)}{\partial\phi_x\partial\phi_y}
 \longrightarrow \nonumber \\
\frac{\exp \left(\frac{m^2}{2}\log \frac{m^2-1}{m^2+1}\right)}
{(m^4-1)\log^{1/2}\frac{m^2+1}{m^2-1}} \sim 1/\sqrt{(m^{2}-1)}\log^{1/2}
\frac{m^{2}+1}{m^{2}-1}
\label{5.4}
\eq
 in agreement  (up to the logarithmic factor)
 with the $m^{2} \longrightarrow 1$ limit of
  $exact$ formula (\ref{11}) for the partition
function. This is remarkable, since there is no $a~priori$ reason to
rely upon applicability of the mean--field approximation for $N=2$.
This opens the possibility to believe that analogous results can have
something to do with reality at $d>1$ as well.

The explanation of the picture Fig.3b is in fact very simple and seems
essentially independent of $N$. From formula (\ref{IZ1}) it is
clear that for infinitely large and $different$ values of the
eigenvalues $\phi_i$ the potential is grows quadratically as long as
$m^{2} > m^2_{\rm crit}=d$.  When two eigenvalues come close
to each other, the potential grows logarithmically and they repel.
This is because of the obvious
singularity, arising from $\Delta(\phi)$ in the denominator in (\ref{IZ1})
(and could be interpreted as $attraction$ of eigenvalues for $d>1$) is
in fact overcome by the zero (of the order $2d$) in the sum in
brackets in (\ref{IZ1}) (or the numerator in (\ref{IZ}). It looks very
plausible that this behaviour, which is seen clearly in the case $N=2$,
shown in the Fig.3b (repulsion at coinciding $\phi_i$'s and
quadratic well at infinity) should persist for all N.

However, the properties of this minimum are drastically different at
$d=1$ and at $d>1$. Indeed, at $d=1$ the value of the master field
$\varphi_0$ tends to infinity as one approaches the critical point
$m^2_{\rm crit} = d$ from above, while $\partial^2
V(\varphi_0)/\partial\varphi^2$ tends to zero, thus giving rise to
singularities and the critical behaviour of the partition function,
but this is no longer true if $d>1$. The logarithmic term $+(d-1)\log
\phi^2$ in the effective potential (\ref{pot}) forces it to grow at
large values of $\phi^2$ even when $m^2=d$. Thus the position of the
minimum remains finite, the curvature $\partial^2 V(\varphi_0)
/\partial\varphi^2$ at the minimum does not vanish even for infinitely
small $m^2 - d$ and no critical behaviour occurs. This makes the
result of ref.\cite{G} very natural.  The earlier claim of
ref.\cite{M} that higher order terms in the $\Phi$-potential can be
adjusted to produce critical behavior means  that the coefficient of
the ${\rm tr}\Phi^4$ term in the potential can be
adjusted to make the curvature small.

This, however, does not seem to be the $simplest$ possibility to
overcome the problem, found in \cite{G}. Our interpretation of the
loss of criticality at $d>1$ attributes it to a $logarithmic$ term in
the effective potential (which arises from the Van-der-Monde
determinants $\Delta(\phi)$. This effect can easily be compensated by
adding a $logarithmic$ term to the $bare$ potential. This is
equivalent to introducing $({\rm det}\Phi)^{\alpha}$ in the measure of
integration of original KMM (\ref{1}):
\bq
\tilde{\cal Z}_D\equiv
\int d\Phi(x)[dU(x,y)]({\rm det}\Phi(x))^{\alpha}
\exp \left(-\sum_x{\rm tr}V(\Phi(x))~ +
\right.
\nonumber \\
\left.
+~\sum_{<x,y>} {\rm tr} \Phi(x)U(x,y)\Phi(y)U^{\dagger}(x,y) \right).
\label{1mod}
\eq
Clearly $\alpha$ should be non-negative, $\alpha\geq 0$, in order to
leave the integral over Hermitean matrices $\Phi$ well defined. This
is exactly what necessary in order to restore the critical behaviour
(for $m^2 \rightarrow d+0$ and $d>1$): we just need $\alpha$ to be
positive (and big enough). In fact for $N=2$ the $shape$ of the
effective potential in Fig.3b remains the same for $\alpha \neq 0$,
only the increase at the origin is a bit faster: $-\log\phi^2
\rightarrow -(\alpha +1)\log\phi^2$, while the asymptotics at infinity
$(d-1)\log\phi^2 + (m^2-d)\phi^2 \rightarrow (d-1-\alpha)\log\phi^2 +
(m^2-d)\phi^2$, is clearly dominated by the $\phi^2$ term if $\alpha
\geq d-1$. For $N>2$ the picture becomes more complicated, since ${\rm
det}\Phi$ and $\Delta(\Phi)$ are very different functions of the
eigenvalues of $\Phi$.  (Also note, that the trace part of $\Phi$ does
not decouple as $\alpha \neq 0$.) However, for big enough $\alpha$
there usually will be critical behaviour at some minima with large and
non-coinciding $\phi_i$'s.

Indeed, one can think about the entire set of eigenvalues as of a
system of $N$ point particles on a line (a variant of the Dyson gas).
These particles interact through a logarithmic repulsion at short
distances and a logarithmic attraction at long distances $(d-1)
\log(\phi_i-\phi_j)^2$: a rather peculiar picture of a
Coulomb gas with a logarithmic ``hard core'' which is also
Coulomb-like. The particles also interact with the origin through a
combination of a central harmonic potential
 and a logarithmic repulsion, $(m^2-d)\phi_i^2 -
\frac{\alpha}{2}\log\phi_i^2$. Furthermore,
 since the matrices are traceless, the center of mass of the particles is
constrained to be
at the origin. (Actually, we could consider matrices with a non-zero trace but,
as we have discussed previously, the trace is an irrelevant degree of freedom.)

As long as $m^2-d>0$ the longest range part of the interaction
 is harmonic and
the particles are confined in the vicinity of
 the origin with
size of the order $1/\sqrt{m^2-d}$.  When $m^2-d=0$,
the particles still have a Coulombic long-ranged attraction and if $\alpha $
is not too large, they still form a stable cloud.   When $\alpha $ is large
enough, the repulsive central potential destabilizes the cloud and repels
the particles to infinity.  This is the appropriate critical behavior. This
system is a rather
simple object both for gedanken experiments and computer simulations;
moreover, as we tried to demonstrate, the most interesting qualitative
results can be extracted from the study of the simplest cases,
including $N=2$.

The simple analogue of the model (\ref{1mod}),
\bq
\int d\Phi ({\rm det}\Phi)^{\alpha} e^{-{\rm tr}V(\Phi) +
{\rm tr}\Lambda\Phi},
\label{GKM}
\eq
with a single $\Phi$-matrix and a slightly different interaction with
the matrix-valued background field $\Lambda$ (analogue of $U$), which
is known as the Generalized Kontsevich Model \cite{GKM}, have been
analyzed in \cite{CheMa}. At least for this model the effect of
non-vanishing $\alpha$ is known to create no problems for the
solvability of the model, and is, moreover, natural from the view of
its integrability structure. Introduction of the parameter $\alpha$
even proved useful for description of continuum limits of the model
\cite{CheMa}. All this can easily appear true for the modified KMM
(\ref{1mod}), where the dependence on the potential $V(\Phi)$ should be
studied at least with the purely theoretical motivations (to reveal
the intrinsic integrable structure, for example). This, however, is
the deal for the future.

\section{Conclusion}

This paper was devoted to analysis of the $\Phi$-sector of the
``Gaussian'' KMM. The really interesting properties of the KMM (see
\cite{KMSW}) are essentially independent of the structure of the
$\Phi$-sector.  It is only necessary that it guarantees the existence
of $some$ master-field $\phi$ and that the magnitude of the master
field is $large$. We saw that these requirements were fulfilled in
$all$ of the situations we discussed above, even if the critical point
has nothing to do with the continuum limit.

It is of course important to understand the variety of possible
universality classes associated with the critical behavior.  This is
why it is also interesting to investigate the $\Phi$-sector of the
model.  A priori in the KMM it is not forbidden to look for a
$non-trivial$ universality class represented by a rather simple
``Gaussian'' model.

We showed that this model indeed exhibits a rich pattern of critical
behaviour even at $d=1$, but, as we demonstrated, only the most naive
large-$N$ limit has anything to do with the $continuum$ limit: other
types of the critical points (defined as the places in parameter space
where the partition function is non-analytic) do not survive as the
volume of the system become infinite (they are infrared unstable
critical points). Thus, unconventional large $N$ limits cannot save
the $d>1$ Gaussian KMM from the conclusion of \cite{G}.

Therefore, it is necessary to adopt another line of reasoning.  In
Section 5 we proposed a simple explanation of the result of \cite{G}:
in the Gaussian model and when $d>1$ the effective potential for the
eigenvalues of $\Phi$ remains convex for all $N$.  This was due to a
logarithmic attraction of the eigenvalues at large distances.  We can
estimate the magnitude of the eigenvalues at their equilibrium
positions and it is not so large, probably, of order 1-10, probably
not large enough for the mean field approximation to be effective for
the $U$-sector of the theory.

This reasoning also suggests a way to overcome this difficulty.  We
introduce a repulsive logarithmic central potential which increases
the magnitudes of the eigenvalues at their equilibrium positions and,
if it is strong enough and when the curvature of the Gaussian
vanishes, it gives a critical behavior where some or all of the
eigenvalues take on infinite (or at least arbitrarily large)
magnitudes.  This logarithmic potential can be regarded as a
modification of the integration measure in the partition function of
the Gaussian KMM.  This is a rather mild modification of the model and
may not affect its chances to be exactly solvable.

\vskip  0.5in
{\bf Acknowledgments}

A.M. acknowledges the hospitality of the theory group of the Physics
Department at UBC.

\newpage
\vskip .3in
\bf{Figure Captions:}
\vskip .3in

\noindent
Fig. 1a.  Pictorial representation of the trace of a product of
$\Phi$-matrices.
Dots correspond to the matrices and labels $a_1,\ldots$ correspond
to pairs of indices in the matrix product.

\medskip
\noindent
Fig. 1b. Using Wick's theorem to compute the correlation
functions we must
consider contractions of the matrices, the first kind of which we
denote by a solid line and is
depicted here.  This contraction of two matrices identifies the the
indices as shown.

\medskip
\noindent
Fig. 1c. The other contraction, denoted by a wavy line,
identifies the indices
as depicted here.

\medskip
\noindent
Fig. 2  ~~ The boquet.

\medskip
\noindent
Fig. 3a ~~ The effective potential for SU(2) when
$m^2<m_{\rm crit}^2$ when $d>1$.

\medskip
\noindent
Fig. 3b ~~  The effective potential for SU(2) when
$m^2\geq m_{\rm crit}^2$
when $d>1$.

\end{document}